\documentclass[conference,a4paper]{IEEEtran}

\usepackage{graphicx}
\usepackage{bbm}
\usepackage{algorithmic}
\usepackage{algorithm} 
\usepackage{theorem}
\usepackage{amssymb}
\usepackage{subfigure}
\theoremheaderfont{\upshape\bfseries}
\theorembodyfont{\itshape}
\theoremstyle{plain}

\newtheorem{Lem}{Lemma}

\newtheorem{Def}{Definition}
\newtheorem{Theorem}{Theorem}

\begin{document}

\sloppy

\title{Coverage by Pairwise Base Station Cooperation under Adaptive Geometric Policies}

\author{
  \IEEEauthorblockN{Fran\c{c}ois Baccelli}
  \IEEEauthorblockA{Simons Chair in Mathematics and ECE\\
  UT Austin, USA \&       INRIA-ENS, France\\
    baccelli@math.utexas.edu}
  \and
  \IEEEauthorblockN{Anastasios Giovanidis}
  \IEEEauthorblockA{INRIA-ENS\\
	Paris, France\\
    giovanid@enst.fr}
}

\IEEEspecialpapernotice{(Invited Paper)}


\maketitle

\begin{abstract}

We study a cooperation model where the positions of base stations follow a Poisson point process distribution and where Voronoi
cells define the planar areas associated with them. For the service of each user, either one or two base stations are involved. If
two, these cooperate by exchange of user data and reduced channel information (channel phase, second neighbour interference) with conferencing over some backhaul link. The total user transmission power is
split between them and a common message is encoded, which is coherently transmitted by the stations. The decision for a user to choose service with or without cooperation is
directed by a family of geometric policies. The suggested policies further control the shape of coverage contours in
favor of cell-edge areas. Analytic expressions based on stochastic geometry are derived for the coverage probability in the network. Their numerical evaluation shows benefits from cooperation, which are enhanced when Dirty Paper Coding is applied to eliminate the second neighbour interference.

\end{abstract}

\section{Introduction}

Cooperation in cellular networks has been recently suggested as a promising scheme to improve system performance, especially for cell-edge users \cite{GesbertTutor}, \cite{JorswieckNOW13}. In this work, we study a cooperation model where the positions of Base Stations (BSs) follow a Poisson point process (p.p.p.) distribution and where Voronoi cells define the planar  
areas associated with them. The approach is known to provide closed (or integral) form expressions for important performance metrics averaged out over such large random networks. 

We consider here a specific scheme of cooperation with coherent transmission, where at most two BSs can serve a single user. The scheme is based on the ideas of conferencing and user-power splitting. It originates from the seminal work of Willems \cite{WillemsConf83} for the $2\times 1$ cooperative Multiple Access Channel (MAC) and its Gaussian variant in \cite{BrossLapWigg08}. An adaptation of the idea in a network-MIMO setting is proposed by Zakhour and Gesbert for the $2\times2$ model in \cite{ZakhourGesbSP11}, where the beamforming vectors take the role of signal weights and define the power-split ratios. Our problem position within the framework of stochastic geometry, extends previous work by Andrews, Baccelli and Ganti \cite{AndrewsCoverage} to include cooperative schemes.

More specifically, for the service of each user, either one or two BSs are involved (Section \ref{SII}). If two, these cooperate by exchange of information with conferencing over some backhaul link. We assume that, in addition to the user data, only part of the total channel state is available and exchanged. We choose to analyze schemes with partial knowledge in order to investigate possible benefits of cooperation without the excessive costs from full channel adaptation. Specifically, only the phase of the channel from the second closest transmitter to the user is known, while the first BS is informed over this and transmits coherently by appropriate choice of its own phase. The scheme considers a fixed transmission power budget per user - which is split between the two BSs and a common message is encoded. The two transmitted common signals add up in-phase at the user receiver, resulting in an extra term for the beneficial signal (Section \ref{SIII}). Later in the work, extra knowledge of the interference from the second BS neighbour to the reference user is considered available. The cooperative pair then transmits orthogonaly to this signal by application of Dirty Paper Coding (Section \ref{SIVd}).

Each user chooses on its own to be served with or without BS cooperation. The decision is directed by a family of geometric policies, which depend on the ratio of the user distance to its first and second closest geographic BS neighbour and some design parameter $\rho$, left as optimization variable (Section \ref{SIV}). In this way the plane is divided into non-overlapping zones of cooperation ($\mathrm{Full\ Coop}$) and no cooperation ($\mathrm{No\ Coop}$).

An exact expression of the coverage probability in the network under study is derived (Section \ref{SV}). Numerical evaluation allows one to analyze coverage benefits compared to the non-cooperative case. These benefits are more important with Dirty Paper Coding (Section \ref{SVI}). It is concluded (Section \ref{SVII}) that cooperation can significantly improve coverage without exploitation of further network resources (frequency, time, power), even with schemes of reduced channel state information. 

This work continues the line of research on applications of point processes to wireless networks and takes a step further to consider cooperative transmission in cellular networks. During the last years important results have been derived by use of stochastic geometry tools. The theoretical aspects of point processes and its applications in telecommunications can be found in the book of Baccelli and B\l aszczyszyn \cite{BaccelliBookStoch}. Important contributions on the capacity of wireless networks appear in the book of Weber and Andrews in \cite{WeberAndrewsNOW12} and for K-tier networks in the work of Dhillon et al \cite{DhillonKTIER}, as well as Keeler, B\l aszczyszyn and Karray  in \cite{KeelerBartekISIT13}. Haenggi and Ganti have investigated the modeling and analysis of network interference by use of point processes \cite{HaenggiNOW09}. In the problem of cooperation, parallel to our work Akoum and Heath \cite{AkoumHeathJou13} have provided approximative performance evaluation when BSs are randomly grouped into cooperative clusters. The transmission scheme they consider is intercell interference nulling. Their approach differs from our work, because we consider optimal geographic association of each user with a pair of BSs and coherent transmission (after conferencing) from the latter. All proofs of the analysis in this paper can be found in our extended version \cite{AGFBarXiv13}.

\section{Geometry of Cooperation}
\label{SII}

For the model under study the BSs considered are equipped with a single antenna and are positioned at the 
locations of atoms from the realization of a planar p.p.p. with intensity $\lambda$, denoted by $\phi=\left\{\mathbf{z}_i\right\}$. 
A planar tessellation, called the 1-Voronoi diagram \cite{CompGeomBook}, partitions (up to Lebesgue measure zero) the plane into subregions called cells. The \textbf{1-Voronoi cell} $\mathcal{V}_1\left(\mathbf{z}_i\right)$ associated with $\mathbf{z}_i$ is the locus of all points in $\mathbb{R}^2$ which are closer to $\mathbf{z}_i$ 
than to any other atom of $\phi$. We consider Euclidean distance $ d\left(\mathbf{z}_i,\mathbf{z}\right)$. Then 

\begin{eqnarray}
\mathcal{V}_1\left(\mathbf{z}_i\right)  = \left\{\mathbf{z}\in\mathbb{R}^2|\ d\left(\mathbf{z}_i,\mathbf{z}\right)\leq d\left(\mathbf{z}_j,\mathbf{z}\right), \forall \mathbf{z}_j\in\phi\setminus\left\{\mathbf{z}_i\right\}\right\}.\hspace{-0.25cm}
\end{eqnarray} 
%
When the 1-Voronoi cells of two atoms share a common edge, they constitute \textbf{Delaunay neighbours} \cite{CompGeomBook}. A dual graph of the 1-Voronoi tessellation called the Delaunay graph is constructed  if Delaunay neighbours are connected by an edge. The 2-Voronoi diagram \cite{CompGeomBook} consitutes another partition (up to Lebesgue measure zero) of the plane. Specifically, the \textbf{2-Voronoi cell} $\mathcal{V}_2\left(\mathbf{z}_i,\mathbf{z}_j\right)$ associated with $\mathbf{z}_i,\mathbf{z}_j\in\phi$, $i\neq j$ is the locus of all points in $\mathbb{R}^2$ closer to $\left\{\mathbf{z}_i,\mathbf{z}_j\right\}$ than to any other atom of $\phi$, i.e.
\begin{eqnarray}
\mathcal{V}_2\left(\mathbf{z}_i,\mathbf{z}_j\right) & = & \left\{\mathbf{z}\in\mathbb{R}^2|\right. d\left(\mathbf{z}_i,\mathbf{z}\right)\leq d\left(\mathbf{z}_k,\mathbf{z}\right)\ \& \nonumber\\ 
& & d\left(\mathbf{z}_j,\mathbf{z}\right)\leq d\left(\mathbf{z}_k,\mathbf{z}\right), \left.\forall \mathbf{z}_k\in\phi\setminus\left\{\mathbf{z}_i,\mathbf{z}_j\right\}\right\}. \hspace{+0.3cm}
\end{eqnarray}
The 1-Voronoi tesselation for $10$ randomly positioned atoms is shown in Fig. \ref{fig:ExampleCoop04} and \ref{fig:ExampleCoop09}. In both, the coloured area depicts a 2-Voronoi example region. In the present work we consider a geometric cooperation scenario based on the following assumptions:

\begin{itemize}
\item Each BS is connected via backhaul links of infinite capacity 
with all its Delaunay neighbours.

\item Exactly one user with a single antenna is initially associated with every BS. Each user is 
located randomly at some point within the 1-Voronoi cell of its BS 
and we write $\mathbf{u}_i\in\mathcal{V}_1(\mathbf{z}_i)$. 

\item Each user $\mathbf{u}_i$ may be served by either one or two BSs. If two, these correspond to the atoms of $\phi$ which are its \textbf{first} and \textbf{second closest geographic neighbour}. If one, it is just the first closest neighbour. We use the notation $\mathbf{b}_{i1}$ and $\mathbf{b}_{i2}$ when referring to these neighbours. By definition the user belongs to $\mathbf{u}_i\in\mathcal{V}_2\left(\mathbf{b}_{i1},\mathbf{b}_{i2}\right)$. 

\item From the point of view of a BS located at $\mathbf{z}_i$, we refer to the user in its 1-Voronoi cell as the \textbf{primary user} $\mathbf{u}_i$ and to all other users served by it but located outside the cell as the \textbf{secondary users}. These constitute a set $\mathcal{N}^s\left(\mathbf{z}_i\right)$, with cardinality that ranges between zero and the number of Delaunay neighbours, depending on the users' position relative to $\mathbf{z}_i$.
\end{itemize}
 
The distance between user $\mathbf{u}_i$ and its first (resp. second) closest BS neighbour 
equals $d\left(\mathbf{b}_{i1},\mathbf{u}_i\right) = r_{i1}$ ($d\left(\mathbf{b}_{i2},\mathbf{u}_i\right) = r_{i2}\geq r_{i1}$). The second nearest BS $\mathbf{b}_2$ can only be one of the Delaunay neighbours of $\mathbf{b}_1$.


\section{Cooperative Pairwise Transmission}
\label{SIII}

The communications scenario in this work applies the following idea. When two BSs cooperatively serve a user in the downlink, its signal is split into one common part served by both, and private parts served by each one of the involved BSs. The common part contains information shared by both transmitters after communication between them over a reliable conferencing link.  

More specifically, for each primary user $\mathbf{u}_i$ located within the 1-Voronoi cell of atom $\mathbf{z}_i$, consider a signal $s_i\in\mathbb{C}$ to be transmitted. The user signals are independent realizations of some random process with power $\mathbb{E}\left[\left|s_i\right|^2\right] = p>0$ and are uncorrelated from other user-signals meaning $\mathbb{E}\left[s_i s_j^*\right] = 0$, $\forall j\neq i$. The signal for user $\mathbf{u}_i$ in the downlink is split in two parts: 
$s_i  =  s_{i}^{(pr)}+ s_{i}^{(c)}.$

\begin{itemize}
\item A \textbf{private (pr)} part sent to $\mathbf{u}_i$ from its first BS neighbour $\mathbf{b}_{i1}:=\mathbf{z}_i$, denoted by $s_{i}^{(pr)}$. The second neighbour does not have a private part to send.
\item A \textbf{common (c)} part served by both $\mathbf{b}_{i1}$ and $\mathbf{b}_{i2}$, which is denoted by $s_{i}^{(c)}$. This part is communicated to both BSs over the backhaul links. For the sake of clarity, we will use the notation
$s_{i}^{(c1)}$, $s_{i}^{(c2)}$ for the common signal transmitted from $\mathbf{b}_{i1}$ and $\mathbf{b}_{i2}$ respectively, although the two signals 
are actually scaled versions of each other.
\end{itemize}
The two parts are uncorrelated random variables (r.v.'s), in other words $\mathbb{E}\left[s_{i}^{(pr)} {s_{i}^{(c)}}^*\right]=0$. Considering power issues, we put the constraint that the total power transmitted from all BSs to serve user $\mathbf{u}_i$ sums up to $p$ to guarantee \textbf{power conservation}. We assume that 
the common part is transmitted by both BSs with the same power percentage $a_i\in\left[0,\frac{1}{2}\right]$, named the \textbf{power-split ratio}.
\begin{eqnarray}
\label{powerconservA}
\mathbb{E}\left[\left|s_{i}^{(pr)}\right|^2\right] = \left(1-2a_i\right)p,\mathbb{E}\left[\left|s_{i}^{(c1)}\right|^2\right] = \mathbb{E}\left[\left|s_{i}^{(c2)}\right|^2\right] =a_ip.\nonumber
\end{eqnarray}

Each BS $\mathbf{z}_i$ transmits a total signal $x_i$. By applying superposition coding, this signal consists of the private and common part for its primary user $\left(s_{i}^{(pr)} + s_{i}^{(c1)}\right)$ and the common parts for all its secondary users $s_{k}^{(c2)}$, $k\in\mathcal{N}^s\left(\mathbf{z}_i\right)$. 
\begin{eqnarray}
x_i & = & s_{i}^{(pr)} + s_{i}^{(c1)} + \sum_{k\in\mathcal{N}^s\left(\mathbf{z}_i\right)} s_{k}^{(c2)}.\nonumber
\label{BSsignal}
\end{eqnarray}
%
The BS signals propagate through the wireless link to reach the users. This process degrades the signal power of BS $\mathbf{z}_j$
received at the location of user $\mathbf{u}_i$, by a factor which depends on the distance $d\left(\mathbf{z}_j,\mathbf{u}_i\right)$ and by the power of a \textbf{complex valued} random fading component $e^{i\theta}\sqrt{g}$ ($i:=\sqrt{-1}$ is the \textit{imaginary number}, not to be confused with the \textit{index} $i$ used always as subscript). 
The fading power is an independent realization of a unit-mean exponential random variable $G$ and the phase is an independent realization of a uniform random variable on $\left[0,2\pi\right)$. We denote the total gain from the first (resp. second) neighbour $\mathbf{b}_{i1}$ ($\mathbf{b}_{i2}$) to user $\mathbf{u}_i$ by $h_{i1} := g_{i1} r_{i1}^{-\beta}$ ($h_{i2} := g_{i2} r_{i2}^{-\beta}$) and the total gain from the first (resp. second) neighbour $\mathbf{b}_{j1}$ ($\mathbf{b}_{j2}$) of some other user $\mathbf{u}_j$, to user $\mathbf{u}_i$ by $h_{j1,i} := g_{j1,i} d_{j1,i}^{-\beta}$ ($h_{j2,i} := g_{j2,i} d_{j2,i}^{-\beta}$), with $\beta>2$. The related channel phases are $\theta_{i1}$ ($\theta_{i2}$) and $\theta_{j1,i}$ ($\theta_{j2,i}$). The total signal received at user $\mathbf{u}_i$ is 

\begin{eqnarray}
y_{i} & = & \left(s_i^{(pr)}+s_i^{(c1)}\right)e^{i\theta_{i1}}\sqrt{h_{i1}} +  s_i^{(c2)}e^{i\theta_{i2}}\sqrt{h_{i2}}\nonumber\\
& + & \sum_{\mathbf{u}_j\neq \mathbf{u}_i}f_j + \eta_{i},\nonumber
\label{RecYui}
\end{eqnarray}
where $f_j:=\left(s_j^{(pr)}+s_j^{(c1)}\right)e^{i\theta_{j1,i}}\sqrt{h_{j1,i}} +  s_j^{(c2)}e^{i\theta_{j2,i}}\sqrt{h_{j2,i}}$. The noise is a realization of the r.v. $\eta_{i}\sim \mathcal{N}\left(0,\sigma_{i}^2\right)$, which follows the normal distribution. In the above the signal sum over $\mathbf{u}_j\neq \mathbf{u}_i$ is the interference received by user $\mathbf{u}_i$. The beneficial signal received at the user location is equal to

\begin{eqnarray}
\label{BenSigTheta}
\mathcal{S}_i^{(\theta)}\left(a_i,p\right) =\hspace{+6cm}\nonumber\\
 h_{i1}\left(1-a_i\right)p+h_{i2}a_ip+2a_ip\sqrt{h_{i1}h_{i2}}\cos\left(\theta_{i1}-\theta_{i2}\right),
\end{eqnarray}
and a similar term with the adequate indexing appears for each interference term due to user $\mathbf{u}_j\neq\mathbf{u}_i$. The term with the $\cos\left(\cdot\right)$ is an extra term which is related to the phases of the channels from the first and second neighbour and can be positive or negative depending on the phase difference. By controlling this term we can maximize the received beneficial signal. \textit{If the phase $\theta_{i2}$ is known and communicated to the first neighbour}, the latter can choose to transmit with $\mathbf{\theta_{i1}=\theta_{i2}}$. As a result the extra term is maximized, since $\cos\left(0\right)=1$. The same action cannot be applied to the interference terms. The control affects only the primary user of each BS. The emitted signal for some user is interference for some other with a random fading phase in $\left[0,2\pi\right)$. The expected value of the interference terms is $\mathbb{E}_{\theta}\left[\cos\left(\theta_{j1,i}-\theta_{j2,i}\right)\right]=0$ and the extra term disappears in expectation. With the above observations the $\mathrm{SINR}$ takes the form

\begin{eqnarray}
\label{SINRui2a1}
\mathrm{SINR}_{i}\left(\mathbf{a},p\right) & = & \frac{\mathcal{S}_i\left(a_i,p\right)}{\sigma_i^2 + \mathcal{I}_i\left(\mathbf{a}_{-i},p\right)}\\
\label{SINRui2a2}
\mathcal{S}_i\left(a_i,p\right) & := & h_{i1}\left(1-a_i\right)p + h_{i2}a_ip + 2a_ip\sqrt{h_{i1}h_{i2}}\nonumber\\ \\
\label{SINRui2a3}
\mathcal{I}_i\left(\mathbf{a}_{-i},p\right) & := & \sum_{j\neq i}\mathcal{S}_{j,i}\left(a_j,p\right)\\
\label{SINRui2a4}
\mathcal{S}_{j,i}\left(a_j,p\right) & := & h_{j1,i}\left(1-a_j\right)p+h_{j2,i}a_jp.
\end{eqnarray}



\begin{figure*}[ht!]
\begin{eqnarray}
\label{SINRo2}
\mathrm{SINR}\left(\rho,r_1,r_2\right) = \frac{g_1 r_1^{-\beta}}{\sigma^2+ \mathcal{I}\left(\rho,r_2\right)} \mathbbm{1}_{\left\{r_1\leq \rho r_2\right\}} + \frac{\frac{\left(\sqrt{g_1 r_1^{-\beta}}+\sqrt{g_2 r_2^{-\beta}}\right)^2}{2}}{\sigma^2 + \mathcal{I}\left(\rho,r_2\right)} \mathbbm{1}_{\left\{r_1> \rho r_2\ \&\ r_1\leq r_2\right\}}\hspace{+2cm}\\
\nonumber\\
\label{LTz}
\mathcal{L}_Z\left(s,\mu_1,\mu_2\right) = g\left(s\right)^{-3}\left(-s\sqrt{\frac{1}{\mu_1\mu_2}}\pi +s\sqrt{\frac{1}{\mu_1\mu_2}} \arctan\left(\sqrt{\frac{\mu_1}{\mu_2}}g\left(s\right)\right)+s\sqrt{\frac{1}{\mu_1\mu_2}} \arctan\left(\sqrt{\frac{\mu_2}{\mu_1}}g\left(s\right)\right)+ g\left(s\right)\right)
\end{eqnarray}
\end{figure*}

\section{Geometric Policies with Binary Action Set}
\label{SIV}

In the above $\mathrm{SINR}$ expressions, the vector of power-split ratios $\mathbf{a}$ is not predifined and it gives a 
continuous range of cooperation possibilities. However, taking the first derivative of the beneficial signal in (\ref{SINRui2a2}) over $a_i$, we can find its maximizers by equating to zero, in other words $\frac{\partial \mathcal{S}_i\left(a_i,p\right)}{\partial a_i}=0$. The two maximizers are either $a_i=0$ or $a_i=1/2$, depending on the value of the ratio $\frac{h_{i2}}{h_{i1}}$. We consider that these are the optimal cases for power-splitting when the choices of other users $\mathbf{a}_{-i}$ are held fixed. Having this in mind, we will focus in our work on policies which switch between the two extreme cases. The relevant $\mathrm{SINR}$ expressions below, result by appropriate substitution in (\ref{SINRui2a1})-(\ref{SINRui2a4}).

\begin{itemize}
\item \textbf{No Cooperation} ($\mathrm{No\ Coop}$) for $a_i= 0$
\begin{eqnarray}
\label{NoC}
\mathrm{SINR}_i\left(0,\mathbf{a}_{-i}, p\right) & = & \frac{h_{i1} p}{\sigma_i^2 + \mathcal{I}_i\left(\mathbf{a}_{-i},p\right)}.
\end{eqnarray}
\item \textbf{Full Cooperation} ($\mathrm{Full\ Coop}$) for $a_i= \frac{1}{2}$
\begin{eqnarray}
\mathrm{SINR}_i\left(\frac{1}{2},\mathbf{a}_{-i},p\right) & = & \frac{\frac{\left(\sqrt{h_{i1}}+\sqrt{h_{i2}}\right)^2}{2}p}{\sigma_i^2 + \mathcal{I}_i\left(\mathbf{a}_{-i},p\right)}.
\label{FC}
\end{eqnarray}
\end{itemize}

We identified above that the criterion for choosing between the two cases is the value of the channel power ratio $\frac{h_{i2}}{h_{i1}}$. However, it is usually better for network performance not to change very often the decision on which cooperative clusters serve users. Hence, the ratio of power path losses $\frac{r_{i2}^{-\beta}}{r_{i1}^{-\beta}}$ and consequently the inverse ratio of distances $\frac{r_{i1}}{r_{i2}}$ can better characterize the optimal choice.

\begin{figure}[t!]    
\centering  
\label{fig:VoronoiREG}
	 		\subfigure[Cooperation Regions $\rho=0.4$.]{          
           \includegraphics[trim = 28mm 55mm 25mm 55mm, clip, width=0.225\textwidth]{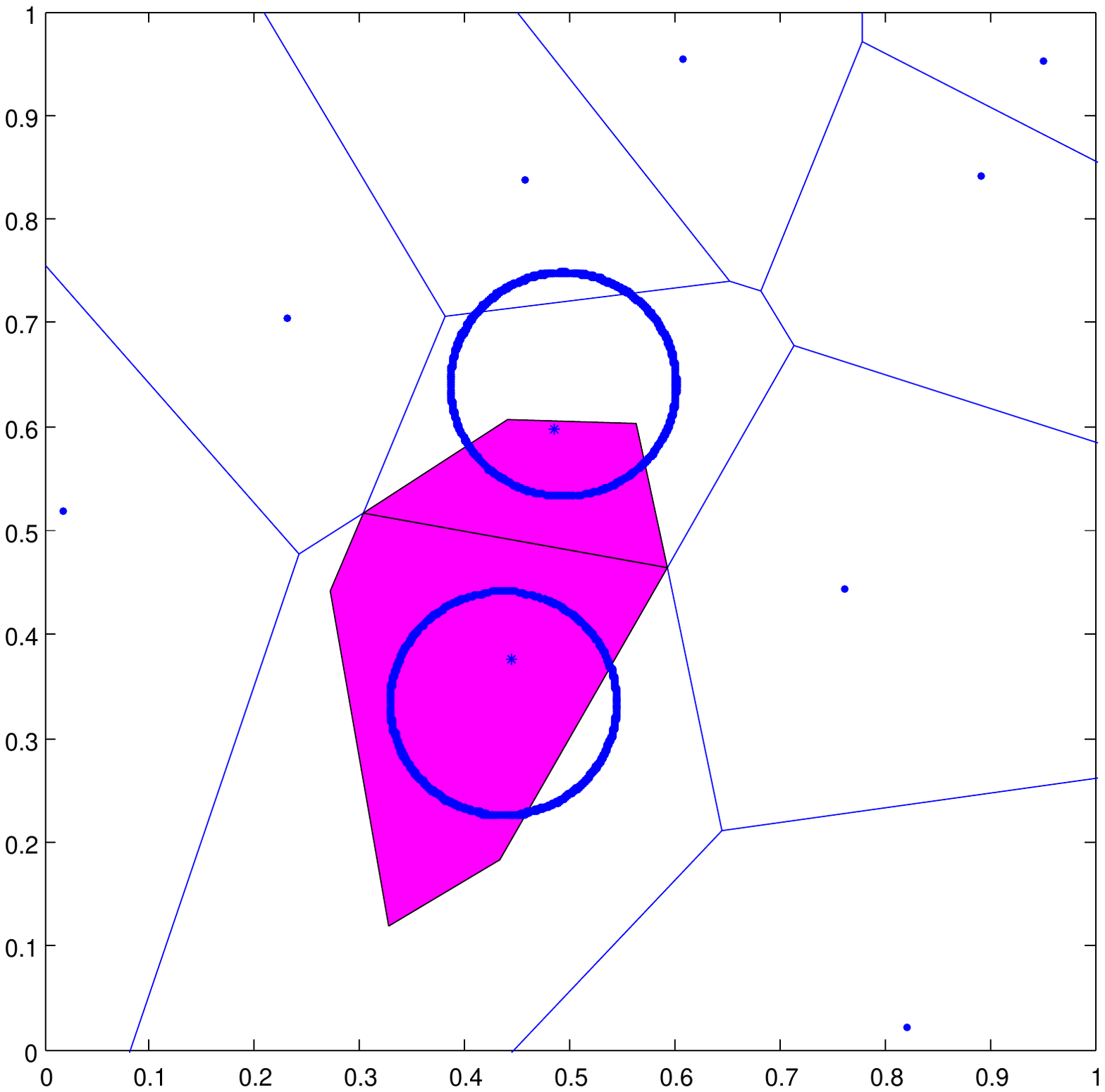}
           \label{fig:ExampleCoop04}
           }
            \subfigure[Cooperation Regions $\rho=0.9$]{          
           \includegraphics[trim = 25mm 55mm 25mm 55mm, clip, width=0.225\textwidth]{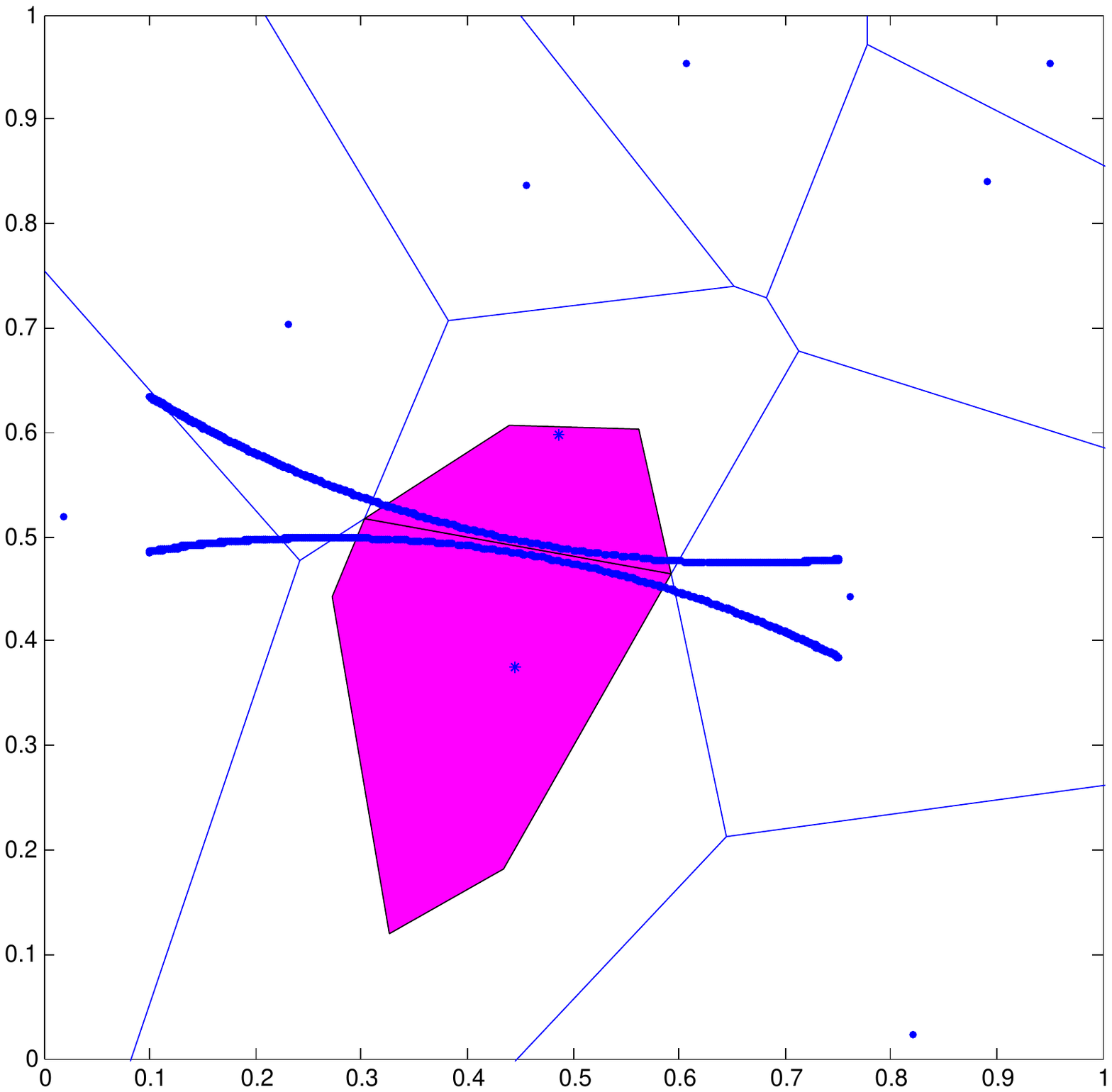}
           \label{fig:ExampleCoop09}
           }
           \caption{Example of 1-Voronoi tesselation in a topology of $10$ uniformly scattered atoms. The coloured area depicts one 2-Voronoi cell and the circles define cooperation regions for its related BS pair.}
\end{figure}

\begin{Def}
\label{PolicyGeomDef}
We consider in this work the family of policies with \textbf{global} parameter $\rho\in\left[0,1\right]$ such that
\begin{eqnarray}
\label{PolicyGeom}
a_i & = & \left\{
\begin{tabular}{l l l}
$0$ & $\left(\mathrm{No\ Coop}\right)$ & , if $r_{i1}\leq \rho r_{i2}$\\
$\frac{1}{2}$ & $\left(\mathrm{Full\ Coop}\right)$ & , if $r_{i1}> \rho r_{i2}$
\end{tabular}
\right..
\label{CaseOPTstoch}
\end{eqnarray}
\end{Def}
The policies are \textbf{user-defined} and \textbf{geometric} because the choice to cooperate or not depends on the relative position of each user $\mathbf{u}_i$ to its two closest BSs. The ratio $\rho\in\left[0,1\right]$ defines the planar cooperation regions, has a global value for the network and is left as optimization variable. Interestingly, the geometric locus of planar points with $r_{i1}\leq \rho r_{i2}$ for $\rho\in\left[0,1\right)$, where the policy chooses $\mathrm{No\ Coop}$ is a \textbf{disc}. For $\rho=1$ the locus degenerates to the line containing the 1-Voronoi boundary of the two cells. 
%
%
%
The two extreme values of $\rho$ give:
\begin{itemize}
\item $\mathrm{Full\ Coop}$ everywhere on the plane when $\rho=0$.
\item $\mathrm{No\ Coop}$ everywhere on the plane when $\rho=1$.
\end{itemize}

Fig. \ref{fig:ExampleCoop04} and Fig. \ref{fig:ExampleCoop09} illustrate two examples for the geometric locus of points of $\mathrm{Full\ Coop}$, given a realization of BS positions. The locus consists of all planar points for which $\mathbbm{1}_{\left\{r_{i1}> \rho r_{i2}\ \&\ r_{i1}\leq r_{i2}\right\}} = \mathbbm{1}_{\left\{r_{i1}> \rho r_{i2}\ \&\ \mathcal{V}_2\left(\mathbf{b}_{i1},\mathbf{b}_{i2}\right)\right\}} =1$, in other words the subregion of the
2-Voronoi cell which lies outside the $\mathrm{No\ Coop}$ discs. The larger the value of $\rho$, the "thinner" the cooperation region.


\section{Stochastic Geometry for Coverage}
\label{SV}

We formulate the problem within the framework of Stochastic Geometry \cite{BaccelliBookStoch}. We focus on a random location on the plane and set its coordinates as 
the Cartesian origin $\left(0,0\right)$. We assume that a user is positioned at this point, whom we denote by $\mathbf{u}_o$. The first (resp. second) closest BS is $\mathbf{b}_1:=\mathbf{b}_{o1}$ ($\mathbf{b}_2:=\mathbf{b}_{o2}$) with distance $r_1=r_{o1}$ ($r_2:=r_{o2}$) from the typical location and 
gains $h_{1}:= h_{o1} = g_{1} r_{1}^{-\beta}$ ($h_{2}:=h_{o2} = g_{2} r_{2}^{-\beta}$), 
$h_{j1}:=h_{j1,o} = g_{j1} d_{j1}^{-\beta}$ ($h_{j2}:=h_{j2,o} = g_{j2} d_{j2}^{-\beta}$).
Furthermore, $g_1$, $g_2$, $g_{j1}$ and $g_{j2}$ are realizations of  independent exponential r.v.'s with mean $p$ (the individual per-user power) $G_{ji},G_i\sim \exp\left(1/p_i\right)$, $p_i=p$, $i=1,2$. Consequently, $\sqrt{G}$ follows the Rayleigh distribution.
 We aim at deriving the \textbf{coverage probability} of the cooperating system

\begin{eqnarray}
q_c\left(T,\lambda,\beta,p,\rho\right)& := & \mathbb{P}\left[\mathrm{SINR}>T\right],
\label{CovDef}
\end{eqnarray}
where the $\mathrm{SINR}$ is measured at a \textbf{typical location} of the plane. Notice that it would be more natural to look at the SINR at the location of a \textbf{typical user} and that the two definitions do not coincide here because the point process of BSs and that of users are not independent. The coverage is a function of the threshold $T$, as well as $\lambda$, $\beta$, $p$ and the policy parameter $\rho$. 
From here on, the parameter set $\left\{T, \lambda, \beta, p\right\}$, which does not influence the analysis, is omitted 
from the arguments. The $\mathrm{SINR}$ for the typical location can now be written using the policies in Def. \ref{PolicyGeomDef} and the expressions (\ref{NoC}) and (\ref{FC}). It takes the form (\ref{SINRo2}). The paragraphs to follow build up the necessary analytical blocks to calculate (\ref{CovDef}).


\subsection{Distribution of Distance to the Two Closest Neighbours}


\begin{Lem}
Given a Poisson p.p. of intensity $\lambda$, the joint p.d.f. of the distances $\left(r_1,r_2\right)$ between the typical location $\mathbf{u}_o$ and its first and second closest neighbour, 
equals
\begin{eqnarray}
f_{r_1,r_2}\left(r_1,r_2\right) & = & \left(2\lambda\pi\right)^2 r_1 r_2e^{-\lambda \pi r_2^2}.
\label{pdfN2f}
\end{eqnarray}
The expected value of the distance $r_2$ equals $\mathbb{E}\left[r_2\right] = \frac{3}{4\sqrt{\lambda}}$.
\label{LemJointD}
\end{Lem}

Using the p.d.f. above and the geometric policies, it can be shown that the parameter $\rho\in\left[0,1\right]$ 
fully determines the probability of a randomly positioned user within some 1-Voronoi cell, to choose $\mathrm{No\ Coop}$ as action.

\begin{Lem}
\label{Lemrho}
The probability of a user not to demand for BS cooperation for its service is equal to
\begin{eqnarray}
\mathbb{P}\left[\mathrm{No\ Coop}\right] = \mathbb{P}\left[r_1\leq \rho r_2\right] = \rho^2.
\end{eqnarray}
\end{Lem}


\subsection{Cooperative Channel Fading Distribution}

We derive the general expression for the r.v. distribution 
of the channel fading, when $\mathrm{Full\ Coop}$ is applied.

\begin{Lem}
\label{LemZ}
Given the r.v.'s $G_i\sim \exp\left(1/p_i\right)$, $i=1,2$, the Laplace Transform (LT) of the r.v. 
\begin{eqnarray}
\label{RVz}
Z_{r_1,r_2} := \left(\sqrt{G_1 r_1^{-\beta}}+\sqrt{G_2 r_2^{-\beta}}\right)^2
\end{eqnarray}
is equal to the expression in (\ref{LTz}), where $g\left(s\right) := \sqrt{1+\left(\frac{1}{\mu_1}+\frac{1}{\mu_2}\right) s}$ and $\mu_i:=r_i^{\beta}/p_i$ for $i=1,2$. The p.d.f. is square integrable and its expectation is equal to \\$\mathbb{E}\left[Z_{r_1,r_2}\right] = \frac{1}{\sqrt{\mu_1\mu_2}}\left(\pi/2 + \frac{\mu_1+\mu_2}{\sqrt{\mu_1\mu_2}} \right)$.
\end{Lem}

We further provide an interesting property of the r.v. $Z_{r,r}/2$ of the fading for $\mathrm{Full\ Coop}$ 
with relation to the r.v. $G$, which is the case for $\mathrm{No\ Coop}$. 
We use the notion of the \textit{Laplace-Stieltjes transform ordering} based on which, the r.v. $Y$ dominates the r.v. $X$ and we write $X\leq_L Y$, if 
$\mathcal{L}_X\left(s\right)\geq \mathcal{L}_Y\left(s\right)$, for all $s\geq 0$.

\begin{Lem}
\label{LaplaceOrd}
Given $\mu_1=\mu_2=\mu:=r^{\beta}/p$ and the two r.v.'s $G\sim\exp\left(\mu\right)$ and $Z_{r,r}/2$ from (\ref{RVz}), it holds 
$G\leq_L Z_{r,r}/2$.
\end{Lem}

\begin{figure*}[ht!]
\begin{eqnarray}
\label{TintAll}
q_c\left(\rho\right)
\label{Tint1}
						& = & 	\int_0^{\infty} \! \int_{\frac{r_1}{\rho}}^{\infty} \!  \left(2\lambda\pi\right)^2 r_1 r_2e^{-\lambda \pi r_2^2} \cdot e^{-\frac{r_1^{\beta}}{p} T\sigma^2}\mathcal{L}_{\mathcal{I}}\left(\frac{r_1^{\beta}}{p} T,\rho,r_2\right)\,  \mathrm{d} r_2 \,  \mathrm{d} r_1\nonumber\\
\label{Tint2}
						& + & 	\int_0^{\infty} \! \int_{r_1}^{\frac{r_1}{\rho}} \!  \left(2\lambda\pi\right)^2 r_1 r_2e^{-\lambda \pi r_2^2} \int_{-\infty}^{\infty} \! e^{-2i\pi\sigma^2 s} \mathcal{L}_{\mathcal{I}}\left(2i\pi s,\rho,r_2\right)\frac{\mathcal{L}_{Z}\left(-i\pi s/T,\frac{r_1^{\beta}}{p},\frac{r_2^{\beta}}{p}\right)-1}{2i\pi s}\,  \mathrm{d} s\,  \mathrm{d} r_2 \,  \mathrm{d} r_1
\end{eqnarray}
\end{figure*}

\subsection{Interference as Shot Noise}
\label{ParagInterference}
We describe the interference $\mathcal{I}\left(\rho,r_2\right)$ as a shot-noise field \cite[Ch.2]{BaccelliBookStoch} 
generated by a point process outside a ball of radius $r_2$. We consider all the power splitting decisions of primary users $\mathbf{u}_j$ (either $a_j=0$ or $a_j=1/2$) related to BSs with distance $d_{j1}\geq r_2$ from the origin. The decisions are determined by the value of the global parameter $\rho$ of the geometric policies. The interference received at $\mathbf{u}_o$ is equal to 
\begin{eqnarray}
\label{INtrfRHO}
\mathcal{I}\left(\rho,r_2\right) = \sum_{\mathbf{u}_j\neq\mathbf{u}_o}h_{j1}\mathbbm{1}_{\left\{r_{j1}\leq \rho r_{j2}\right\}}+\frac{h_{j1}+h_{j2}}{2}\mathbbm{1}_{\left\{r_{j2}\geq r_{j1}>\rho r_{j2}\right\}}.\nonumber
\end{eqnarray}
We associate each BS with a r.v. $B_j\sim \mathrm{Bernoulli(\rho^2)}$ such that (from Lemma \ref{Lemrho})

\begin{eqnarray}
\label{Bernou}
B_j & = & \left\{
\begin{tabular}{l l l}
$1$ & with prob. $\rho^2$ & ($\mathrm{No\ Coop}$)\\
$0$ & with prob. $1-\rho^2$ & ($\mathrm{Full\ Coop}$)
\end{tabular}
\right..
\end{eqnarray}
This r.v. models the randomness of user position within each 1-Voronoi cell, which further determines the action chosen for this user, depending on the ratio $\frac{r_{j1}}{r_{j2}}$. 
\begin{itemize}
\item If $B_j=1$, an independent mark $\mathcal{M}_j$ is associated with the BS. The mark is equal to $\mathcal{M}_j:= d_{j}^{-\beta}G_{j}$. 
The signal has to traverse a distance of $d_j=d_{j1}$ from the closest neighbour of user $\mathbf{u}_j$.
\item If $B_j=0$, an independent mark $\mathcal{N}_j$ is associated with the BS. This is the case of full cooperation, where the interfering signal due to user $\mathbf{u}_j$ is coherently emitted from its two closest neighbours. Here, we make the \textbf{far field approximation} $d_{j2}\approx d_{j1} = d_j$, so that the distances of the two cooperating atoms to the typical location are treated as equal. 
Based on this approximation, BSs with primary users requiring $\mathrm{Full\ Coop}$, are considered to emit the entire signal $\mathcal{N}_j := d_j^{-\beta}\frac{\left(G_{j1}+G_{j2}\right)}{2}$, $G_{j1},G_{j2}\sim\exp\left(1/p\right)$.
\end{itemize}
%

The r.v. $G_j$ related to the mark $\mathcal{M}_j$ follows the exponential - or equivalently $\Gamma\left(1,p\right)$ distribution with expected value $p$, whereas the r.v. $\frac{G_{j1}+G_{j2}}{2}$ related to $\mathcal{N}_j$ follows the $\Gamma\left(2,\frac{p}{2}\right)$, with the same expected value $p$. In other words, the path-loss of the interfering signals is in expectation equal in both cases. The interference r.v. is given by

\begin{eqnarray}
\label{Interference1}
\mathcal{I}\left(\rho,r_2\right) := r_{2}^{-\beta} G_{2}B_2 + r_2^{-\beta}\frac{G_1+G_2}{2} \left(1-B_2\right)+ \hspace{+1.5cm}\nonumber\\
+ \sum_{\mathbf{z}_j\in\phi\setminus\left\{\mathbf{b}_1,\mathbf{b}_2\right\}} d_{j}^{-\beta}G_{j} B_j + d_j^{-\beta}\frac{G_{j1}+G_{j2}}{2} \left(1-B_j\right).\nonumber
\end{eqnarray}
The first two terms come from the interference created by the second neighbour lying on the 
boundary of the ball $\mathcal{B}\left(\mathbf{u}_o,r_2\right)$.

\begin{Theorem}
\label{THlaplI}
The LT of the interference r.v. for the model under study, with exponential fast-fading power, is equal to
\begin{eqnarray}
\label{LTinterfD}
\mathcal{L}_{\mathcal{I}}\left(s,\rho,r_2\right) = \mathcal{L}_{\mathcal{J}}\left(s,\rho,r_2\right)e^{-2\pi\lambda\int_{r_2}^{\infty} \!  \left(1-\mathcal{L}_{\mathcal{J}}\left(s,\rho,r\right)\right)r \, \mathrm{d} r },
\end{eqnarray}
where 
\begin{eqnarray}
\label{EachID}
\mathcal{L}_{\mathcal{J}}\left(s,\rho,r\right) = \rho^2\frac{1}{\left(1+sr^{-\beta}p\right)} +\left(1-\rho^2\right)\frac{1}{\left(1+sr^{-\beta}\frac{p}{2}\right)^2}.
\end{eqnarray}
The expected value for the interference r.v. is equal to
\begin{eqnarray}
\label{ExI}
\mathbb{E}\left[\mathcal{I}\left(\rho,r_2 \right)\right] = \frac{p}{\left(\beta-2\right)r_2^{\beta}}\left(\beta-2+2\pi\lambda r_2^2\right),
\end{eqnarray}
and is independent of the parameter $\rho$.
\end{Theorem}




\subsection{Dirty Paper Coding for Second Neighbour Interference}
\label{SIVd}

The second geographic BS neighbour is the dominant factor of interference, due to 
its proximity to the typical location. A mark (either $\mathcal{M}_2$ or $\mathcal{N}_2$) will be strong at the point $\mathbf{u}_o$.
In this subsection we will consider an ideal scenario, where the interference created by the second closest BS can be cancelled out perfectly in the case of full cooperation, by means of coding. This requires precise \textbf{knowledge} by the first neighbour of the interfering signal from the primary user and all possible secondary users served by $\mathbf{b}_2$, which is extra information for the system. If such information is available, the encoding procedure for the typical location can be projected on the signal space of $\mathbf{b}_2$, achievable by Dirty Paper Coding (DPC, see \cite{CostaDP83}) so that the effect of $\mathbf{b}_2$ on interference is eliminated. It the $\mathrm{SINR}$ expression (\ref{SINRo2}) we can then substitute  the variable $\mathcal{I}$ by $\mathcal{I}_{DPC}$, for the case of $\mathrm{Full\ Coop}$. If $\mathbf{u}_o$ chooses $\mathrm{No\ Coop}$, the elimination is not possible. The new r.v. is derived by just omitting the interference part from the second closest BS neighbour.  
Its LT is equal to
\begin{eqnarray}
\label{LTinterfDcancel}
\mathcal{L}_{\mathcal{I}_{DPC}}\left(s,\rho,r_2\right) & = &  e^{-2\pi\lambda\int_{r_2}^{\infty} \!  \left(1-\mathcal{L}_{\mathcal{J}}\left(s,\rho,r\right)\right)r \, \mathrm{d} r },
\end{eqnarray}
where $\mathcal{L}_{\mathcal{J}}\left(s,\rho,r\right)$ is given in (\ref{EachID}). The expected value of $\mathcal{I}_{DPC}$ can be shown to be less than $\mathbb{E}\left[\mathcal{I}\left(\rho,r_2\right)\right]$ in (\ref{ExI}).

\subsection{General Coverage Probability}


\begin{Theorem}
\label{CovProb}
For the cooperation scenario under study and for a given set of system values $\left\{T,\lambda,\beta,p\right\}$,
the coverage probability of a typical location as a function of the parameter $\rho\in\left[0,1\right]$ is equal to (\ref{TintAll}). 

In this expression, $\mathcal{L}_{\mathcal{I}}$ is given in (\ref{LTinterfD}), (\ref{EachID}) and $\mathcal{L}_Z$ in (\ref{LTz}). For the case of Dirty Paper Coding, $\mathcal{L}_{\mathcal{I}}$ should be replaced by $\mathcal{L}_{\mathcal{I}_{DPC}}$ given in (\ref{LTinterfDcancel}).
\end{Theorem}

\section{Numerical Results - Evaluation of Gains}
\label{SVI}

\begin{figure}[t!]    
\centering           
\includegraphics[trim = 5mm 50mm 3mm 50mm, clip, width=0.37\textwidth]{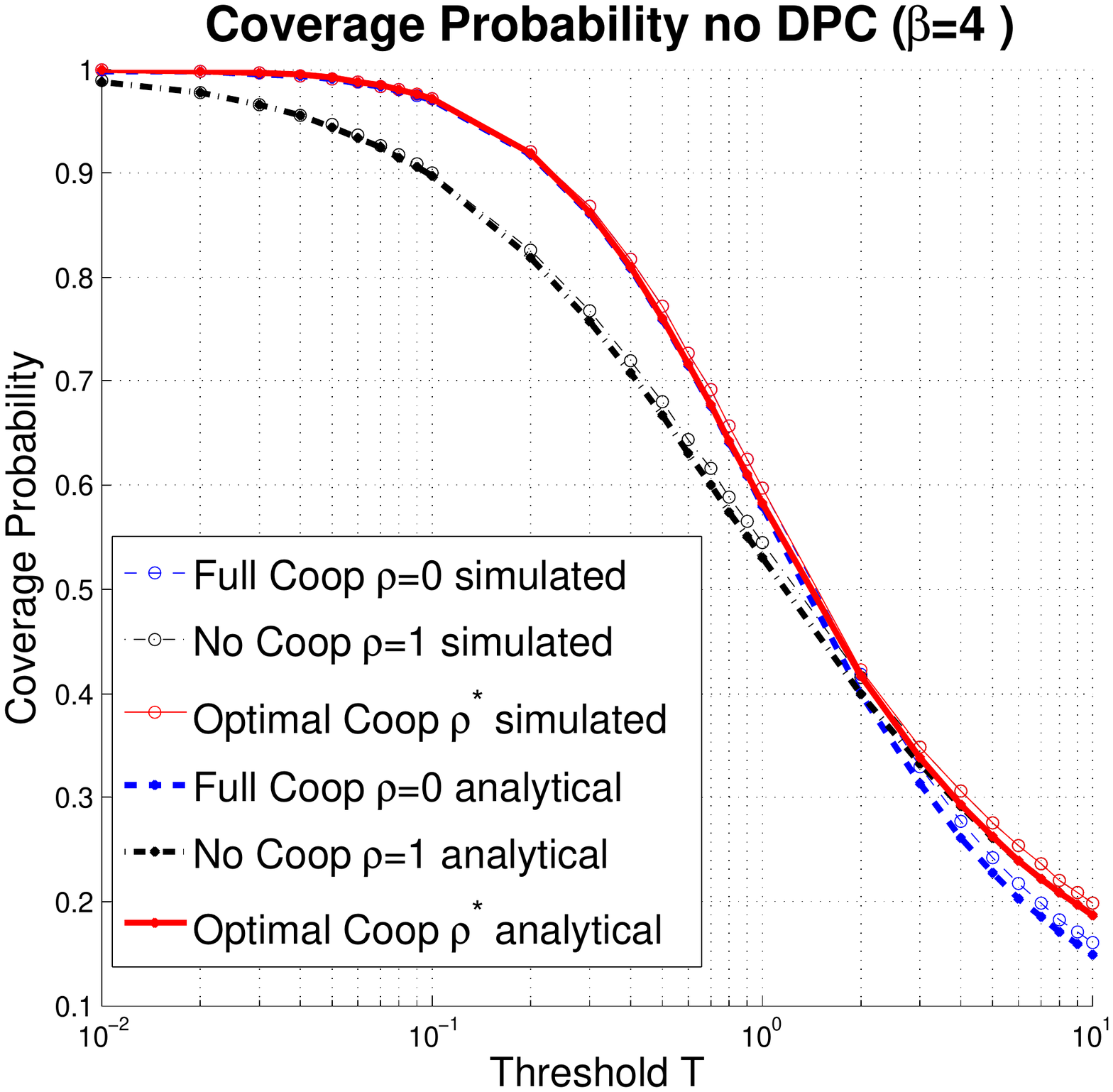}
\caption{Coverage probability without DPC - numerical and simulated}
\label{fig:noDPCF} 
\end{figure}
\begin{figure}[t!]
\centering
\includegraphics[trim = 5mm 50mm 00mm 50mm, clip, width=0.37\textwidth]{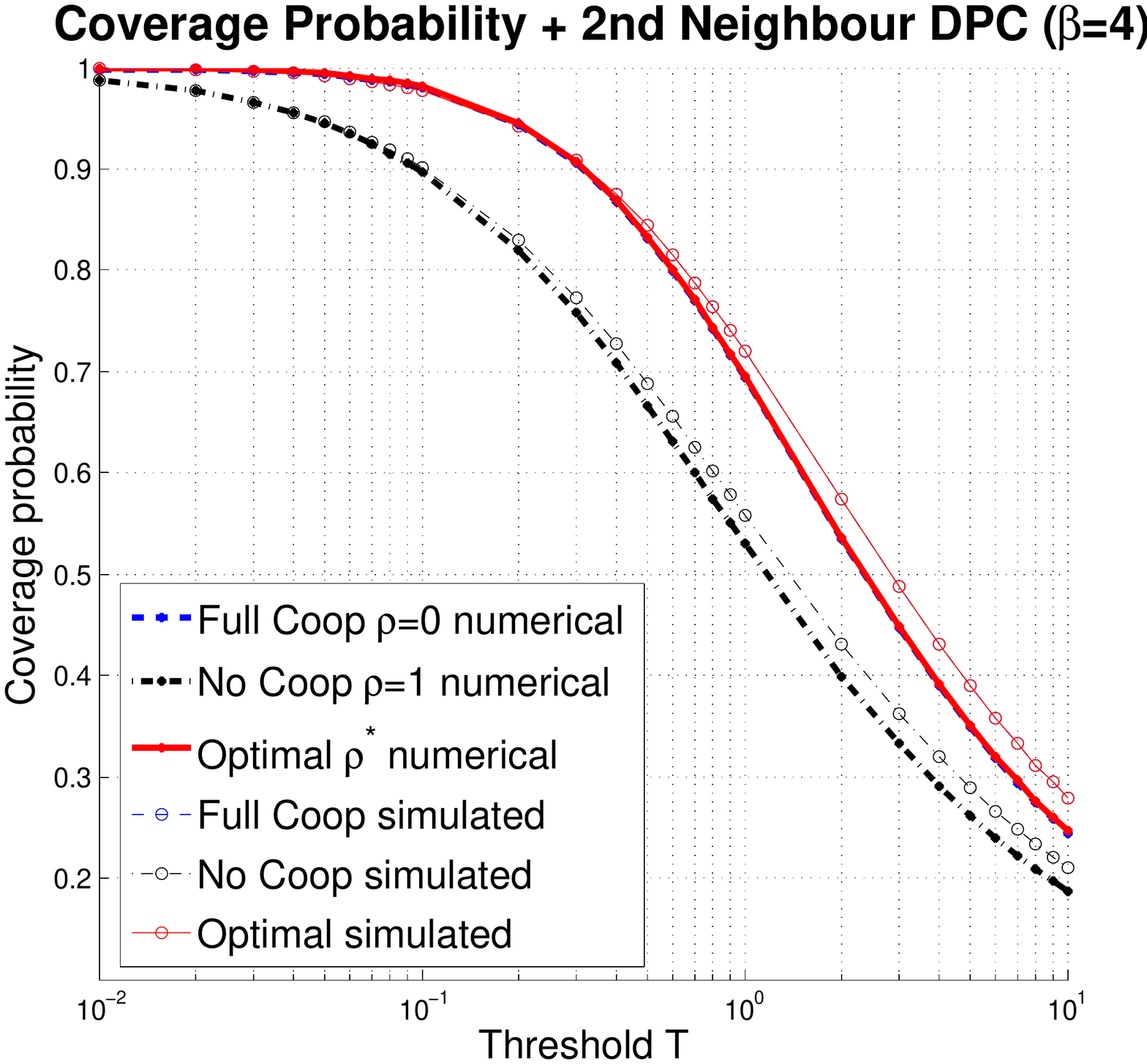}
\caption{Coverage Probability with DPC - numerical and simulated.}
\label{fig:DPC}
\end{figure}

We have numerically evaluated the integrals in (\ref{TintAll}) for parameter density $\lambda=1$, path-loss exponent $\beta=4$, per-user power $p=1$ and noise $\sigma^2=1$. The two cases without and with DPC are shown in Fig. \ref{fig:noDPCF} and Fig. \ref{fig:DPC} respectively. Both figures also include plots from simulations, so that the validity of our approximations can be guaranteed. The simulation results show the coverage probability, taken as an average of 10,000 realizations of different random BS topologies with expected number of atoms $\mathbb{E}\left[N\right]=20$, uniformly positioned in an area of $20\ \mathrm{m^2}$. In each figure, three curves are produced for the numerical evaluation and three for the simulations: (a) coverage with $\mathrm{No\ Coop}$ everywhere $\rho=1$, (b) coverage with $\mathrm{Full\ Coop}$ everywhere $\rho=0$, (c) coverage with optimal $\rho^*$. In (c) the parameter $\rho$ is optimally chosen, so that the coverage $q\left(\rho\right)$ is maximized.

Fig. \ref{fig:noDPCF} illustrates that cooperation with some $\rho^*<1$ can be optimal for $T<2$, whereas $\rho^*=1$ ($\mathrm{No\ Coop}$ everywhere) is always optimal for $T\geq 2$. The maximum gain in coverage with cooperation is $10\%$ between $T=0.1$ to $0.5$. The case of DPC for the interference from $\mathbf{b}_2$ gives more substantial coverage gains, as shown in Fig. \ref{fig:DPC}. The gains appear in the entire domain of $T$ and reach a maximum of $17\%$ for $T=0.2$. 

The figures illustrate the fact that cooperation becomes more beneficial to the network when more information is exploited. Another important quantitative benefit, not visible in the total coverage probability evaluation we show here, is that the coverage area shapes change in favor of cell-edge users while $\rho$ reduces from $1$ to $0$.

%
%

\section{Conclusion}
\label{SVII}

In the present work, we evaluated the coverage probability of a cellular network, when BSs can cooperate pairwise for the service of each user. For this we have applied tools from stochastic geometry. Cooperation is understood here as coherent  transmission of a common message to the user from its two closest BSs. This message is exchanged between them by conferencing, together with the value of the channel phase from the second neighbour. The user can choose between $\mathrm{Full\ Coop}$ or $\mathrm{No\ Coop}$ based on the suggested geometric policies. Closed form expressions for the coverage probability have been derived, whose evaluation quantifies the benefits resulting from cooperation. The benefits are larger when DPC, to avoid second neighbour interference, is applied.

\end{document}